# Length traceability chain based on chromium atom transition frequency

Xiao Deng[1,2,3,4,5], Zichao Lin[1,2,3,4,5], Gaoliang Dai[6,*], Zhaohui Tang[1,2,3,4,5], Zhangning Xie[1,2,3,4,5], Guangxu Xiao[1,2,3,4,5], Zhijun Yin[1,2,3,4,5], Lihua Lei[7], Tao Jin[8], Dongbai Xue[1,2,3,4,5], Zhenjie Gu[1,2,3,4,5], Xinbin Cheng[1,2,3,4,5,*], Tongbao Li[1,2,3,4,5]

[1] Institute of Precision Optical Engineering, Tongji University, Shanghai 200092, China;
[2] MOE Key Laboratory of Advanced Micro-Structured Materials, Tongji University, Shanghai 200092, China
[3] Shanghai Frontiers Science Center of Digital Optics, Tongji University, Shanghai 200092, China
[4] Shanghai Professional Technical Service Platform for Full-Spectrum and High-Performance Optical Thin Film Devices and Applications, Tongji University, Shanghai 200092, China
[5] School of Physics Science and Engineering, Tongji University, Shanghai 200092, China
[6] Physikalisch-Technische Bundesanstalt, Braunschweig, 38116, Germany
[7] Shanghai Institute of Measurement and Testing Technology, Shanghai 201203, China;
[8] University of Shanghai for Science and Technology, Shanghai 200093, China
* Correspondence: gaoliang.dai@ptb.de; chengxb@tongji.edu.cn

## Abstract

Precise positioning measurement plays an important role in in today's advanced manufacturing industry, and length traceability chain has been optimizing and enriching to fulfill the developing and various precise positioning requirement. In this paper, we propose a new length traceability chain based on chromium atom transition frequency, which is a combining utilization of fundamental physical constant's accuracy and grating interferometer's environmental robustness. The self-traceable grating pitch standard, the self-traceable angle standard and the self-traceable grating interferometer are promising to improve the measurement accuracy, consistency and self-calibration ability in situ for precise positioning.

Meter is one of the seven base units of the International System of Units (SI)[1] and the precise positioning measurement, including linear or rotary standard and displacement, plays an important role in in today's advanced manufacturing industry[2,3,4] In order to achieve accurate, reliable and worldwide comparable results for precise positioning, metrological traceability need to be established to guarantee the measurement results can be related to a reference through a documented unbroken chain of calibrations[5,6]. A metrological length traceability chain normally consists of realization of the definition of the Meter, value transfer between the test system and reference system for the measurands, and measurements of displacement and/or rotations[3]. Since references and measurement methods of calibration hierarchy both contribute measurement uncertainty, and the measurement range and accuracy requirements vary in different applications, length traceability chain has been optimizing and enriching to fulfill the developing and various precise positioning requirement[1, 7, 8, 9, 10, 11, 12].

Utilization of fundamental physical constants is the core concept in the realization of definition of the Meter. The International Committee for Weights and Measures (CIPM) issued a recommendation concerning the practical realization of the unit of length[13]: (1) Optical path travelled in vacuum by an electromagnetic plane wave in a time $t$; (2) Wavelength $\lambda$ in vacuum of an electromagnetic plane wave of frequency $v$; (3) Stated wavelength in vacuum or stated frequency of recommended radiations with the specified uncertainty[14], and the most commonly used wavelength is 632.99121258 nm (a relative standard uncertainty of $2.1\times10^{-11}$) of a He-Ne laser with $I_2$ stabilization[15]. Since achieving sub-nanometer accuracy based on wavelength in air requires considerable care of air refractive index correction and nonlinearity terms verification, the lattice parameter of silicon ($d_{220}$=192.0155714×10$^{-12}$ m, with a standard uncertainty of 0.0000032×10$^{-12}$ m) is adopted as a secondary realization of the meter to support dimensional nanometrology[1, 16]. The extremely accuracy of atom transition frequency or the lattice constant lays the basis of measurement accuracy for length traceability chain.

In length traceability chain, optical measurement technologies for precision positioning are key technologies for value transfer between the reference system and test system for the measurands, including laser interferometer, optical encoder/grating interferometer, time-of-flight sensor with a femtosecond laser, laser triangulation sensor and so on[2]. And laser interferometers and optical linear encoder/grating interferometers are two of the major sensor types widely used[17]. The left part of Fig.1 shows the typical length traceability chain based on $I_2$ transition frequency stabilized He-Ne laser with a wavelength of 633nm. Laser interferometry uses the laser wavelength as the scale graduation for measurement, so it holds a direct traceability to the Meter by beating frequency measurement with the primary wavelength standard in National Metrology Institutes. Taking a simple homodyne laser interferometer for example, the light intensity captured by the photodetector can be expressed as

$$I = I_0\left(1+\cos\frac{4\pi n_{air}f\Delta x}{c_0}\right),$$ where $n_{air}$, $c_0$, $f$, $\Delta x$ are the refractive index of air,

the speed of light in a vacuum, the optical frequency, the displacement of a moving target, respectively[2, 18, 19]. Since the refractive index of air can be influenced by temperature, pressure, humidity and so on, the laser interferometer with an un-stabilized He-Ne laser's relative measurement uncertainty around $1\times10^{-6}$ level[3, 20]. For an optical encoder or grating interferometer with a grating pitch of $g$, the typical light intensity captured by the photodetector can be expressed $I = I_0\left(1+\cos\frac{4\pi\Delta x}{g}\right)$, and the grating pitch works as the scale graduation for measurement. Since there is no refractive index of air coupling and the grating is a solid substance, the optical encoder or grating interferometer itself demonstrate good environmental robustness and low noise level[21, 22]. Based on its advantage of much short beam pass and less sensitive to air turbulence, grating interferometers have successfully fulfilled the moving wafer and reticle stage positioning requirements in advanced lithography[23, 24] and become the most widely used sensors for precision positioning in production engineering[3, 25].

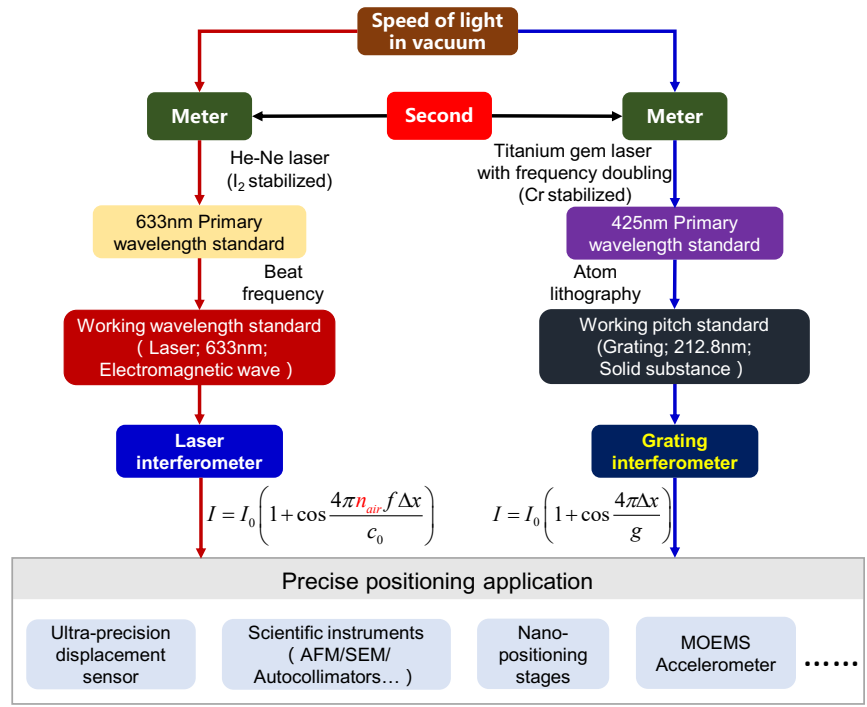

Fig.1 Length traceability chain based on I$_2$ transition frequency and Cr transition frequency

Although the optical encoder or the grating interferometer continue to increase in importance, the lack of direct or independent traceability of the gratings becomes a challenge and limitation. On the one hand, from the prospective of grating manufacturing, the accuracy of the grating pitch directly influences the measurement accuracy of the linear encoders[26, 27]. For the grating fabricated by mechanically scribing, scanning interference lithography or the e-beam lithography, the ruling error, the wavelength drift and nano-positioning error coupled in will lower the grating pitch accuracy. On the other hand, since the grating pitch varies for different samples, the grating pitch and pitch deviation need to be calibrated by metrological methods[2, 22], such as diffraction method based on laser wavelength[28, 29, 30, 31] which in turn introduces errors in the laser wavelength or the laser interferometer. And the periodic calibration and validation of grating pitch is time consuming, requiring skilled labor and expensive reference instrumentation[8]. Meanwhile, there is an increasing need in more higher density gratings for grating interferometers to increase the measurement resolution as the grating pitch works as the scale graduation. Therefore, length traceability chain innovation for optical encoders or grating interferometers will optimize the measurement accuracy, precision and resolution.

Motivated by these aspects above, in this paper, we proposed a new length traceability chain based on chromium atom transition frequency, which is illustrated in the right part of Fig.1. The new length traceability chain utilizes the 425.55nm wavelength stabilized by Cr transition frequency ($^7S_3 \rightarrow ^7P_4^0$) as the realization of Meter. Subsequently, Cr atom lithography technology[32] is adopted to "materialize" the 425.55nm primary length standard in vacuum into solid Cr grating with an exactly period of 212.8nm, the so-called "self-traceable length standard". Then, a self-traceable grating interferometer is proposed to obtain dynamic displacement measurement with direct and independent traceability. Meanwhile, the self-traceable length standard demonstrates the ability of providing self-traceable angle standard and increasing the gating intensity to an extremely high level, i.e., 4700 lines per mm. The length traceability chain is a combining utilization of fundamental physical constant's accuracy and grating interferometer's environmental robustness, which improves the measurement accuracy, consistency and self-calibration ability *in situ* for precise positioning.

## Results

**Self-traceable length standard.** Atom lithography, or the so-called laser focused atomic deposition, utilizes the dipole force to focus the neutral atoms to fabricate high parallel lines. It has been successfully demonstrated using Na[33], Cr[32, 34], Al[35], Yb[36] and Fe[37] Because Cr has low surface diffusion and high chemical stability, the Cr atom lithography is most widely investigated. Fig.2 shows the Self-traceability of Cr grating fabricated by atom lithography[32, 38, 39]. The Cr atom transition frequency adopted here is $^7S_3 \rightarrow ^7P_4^0$, corresponding to a wavelength of 425.55292 nm and a natural linewidth of $\Gamma/2\pi \approx 5MHz$ [39]. When the collimated Cr atoms pass through a near-resonant laser light, grazing across the substrate surface, a dipole force induced by the standing wavelength impose the atom towards the nodes or antinodes of the standing wave, and then the highly parallel gratings forms with a theoretical pitch value of $212.7705 \pm 0.0049$ nm[38]. Since the Cr grating pitch equals to half of the laser wavelength, which is stabilized by Cr atom transition frequency of $^7S_3 \rightarrow ^7P_4^0$, the Cr grating fabricated by atom lithography holds a self-traceability to fundamental natural constant.

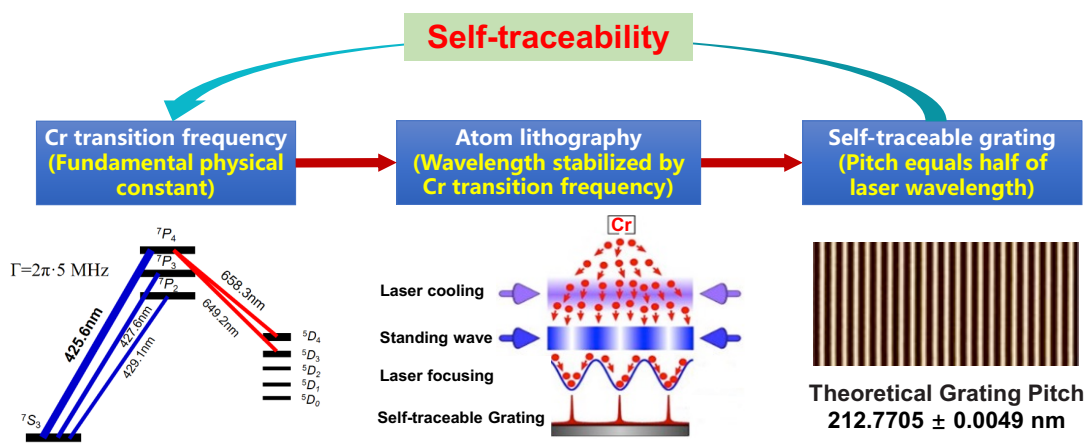

Fig.2 Self-traceability of Cr grating fabricated by atom lithography[32, 38, 39]

In the length traceability chain based on Cr atom transition frequency proposed here, the Cr grating works as a self-traceable length standard, so the accuracy, uniformity, and consistency of the gratings should be carefully examined. Jabez McClelland's group have ever evaluated the accuracy of the Cr grating based on the diffraction method[38], but in fact, the accuracy of grating pitch at trans-scale, the uniformity of the grating pitch, and the pitch accuracy consistency from sample-to-sample have not been analyzed comprehensively yet. And this justification will determine the reliability and reasonability newly proposed length traceability chain. Therefore, we have fabricated the Cr grating and conducted metrological measurements as follows[34, 40]. The Cr power evaporated from an effusion cell, which was heated to 1625 °C. The Cr atom flux passed the stabilization laser, cooling laser, and standing-wave in order. The laser light source system consists of diode-pumped all-solid-state laser Verdi G-12 (532 nm, 11.5 W), continuously adjustable single-frequency titanium gem laser MBR-110 (851.1 nm,1.9 W), and frequency multiplier MBD-200 (425.55 nm, 0.45 W). During the whole Cr grating fabrication process, the vacuum pressure is below $5\times10^{-6}$ Pa and fluorescence frequency stabilization technology is used to lock the laser wavelength of 425.55 nm and a +250 MHz detuning by an AOM is used for the focusing laser of standing wave.

The metrological large range scanning probe microscope (Met.LR-SPM) equipped with ultra-precision laser interferometers at PTB is applied to characterize the Cr grating and the detail description of Met.LR-SPM is in references[41, 42]. The Met.LR-SPM's scanners have not only achieved a resolution of 0.08 nm along the $x$, $y$ and $z$ axes, respectively, but also demonstrate ability of picometer accuracy and uncertainty ability in the international 2D grating comparison of NANO5[43]. The metrological measurements were conducted in the PTB cleanroom center, and the temperature of the Met.LR-SPM and surrounding air was 21.0±0.5 °C, the relative humidity of the air was 42±5 %. Fig.3 shows a measured AFM data on a Cr grating by the Met. LR-AFM. As expected, highly parallel and uniform grating lines are all covered in a 100 μm×5 μm scan. Followed by 5 repeated measurements of a same grating area of 100 μm×10 μm, mean pitches evaluated using the gravity center (GC)[44] method are 212.7790 nm, 212.7802 nm, 212.7796 nm, 212.7796 nm and 212.7790 nm, respectively. The standard deviation of 5 repeated measurements is only 0.0005 nm, which indicates that measurement stability of the Met. LR-AFM is achieved.

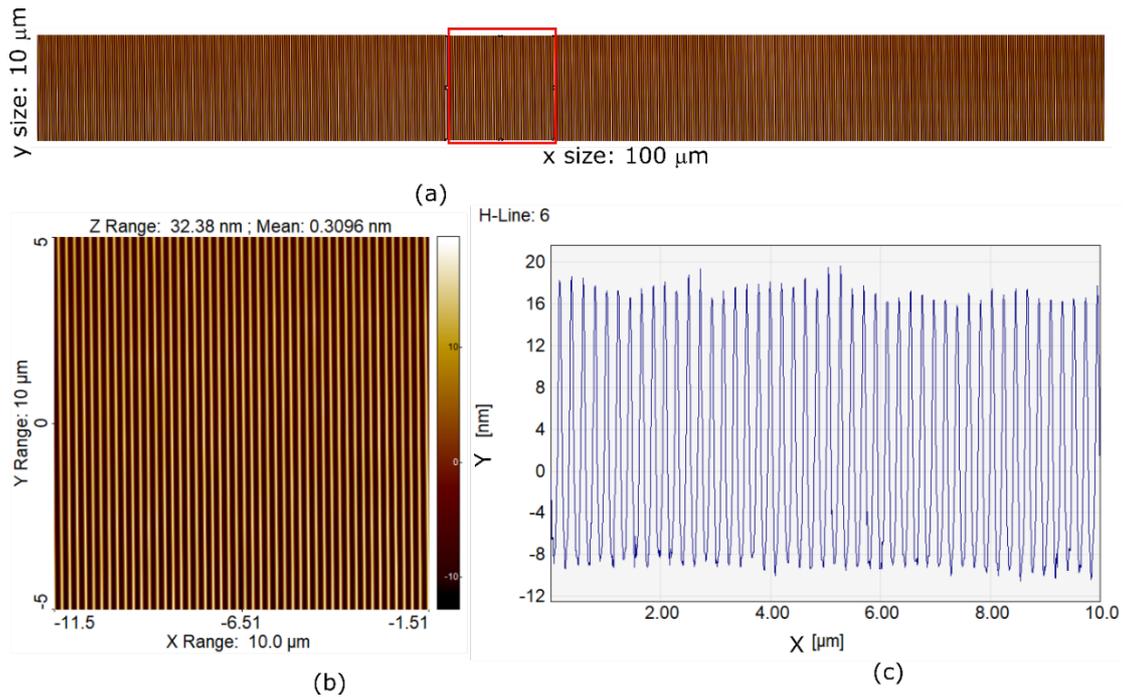

Fig.3 Measured AFM data on a Cr grating by the Met. LR-AFM, shown as (a) an image taken on a feature area with a size of 100 μm x 5 μm and 20000 pixels×10 lines; (b) a detailed view of the features in the marked box in (a); and (c) a cross-sectional profile at the marked line in (b). All data shown are the raw measurement data corrected by the 1st order levelling without filtering and averaging.

Next, we conducted measurements to examine the uniformity of the Cr grating. Table 1 shows averaged mean pitch value calibrated at 20 different measurement locations by Met.LR-SPM, and the data are calculated by GC method and FT-FFT method. Each measurement area has a size of 100 μm×10 μm and is repeatedly measured by 5 times. The mean pitch and standard deviation of different locations calculated by GC and FT-FFT method are 212.7807±0.0011 nm and 212.7805±0.0011 nm. The standard deviation is at 1 picometer level, which indicates extremely good uniformity of the Cr grating.

Table 1 Averaged mean pitch value calibrated at 20 different measurement locations

| No. of measurement Locations | GC Method | | FT-FFT Method | |
|---|---|---|---|---|
| | Averaged mean pitch evaluated using the GC method, nm | Standard deviation, nm | Averaged mean pitch evaluated using the FT-FFT method, nm | Standard deviation, nm |
| 1 | 212.7790 | 0.0005 | 212.7786 | 0.0006 |
| 2 | 212.7783 | 0.0007 | 212.7782 | 0.0004 |
| 3 | 212.7819 | 0.0005 | 212.7816 | 0.0004 |
| 4 | 212.7800 | 0.0003 | 212.7800 | 0.0003 |
| 5 | 212.7812 | 0.0013 | 212.7814 | 0.0013 |
| 6 | 212.7827 | 0.0003 | 212.7829 | 0.0002 |
| 7 | 212.7800 | 0,0004 | 212.7797 | 0.0003 |
| 8 | 212.7801 | 0.0002 | 212.7798 | 0.0001 |

| 9 | 212.7812 | 0.0012 | 212.7809 | 0.0010 |
| 10 | 212,7808 | 0.0010 | 212.7811 | 0.0011 |
| 11 | 212.7804 | 0.0005 | 212.7800 | 0.0005 |
| 12 | 212.7811 | 0.0008 | 212.7808 | 0.0007 |
| 13 | 212.7808 | 0.0015 | 212.7803 | 0.0014 |
| 14 | 212.7820 | 0.0004 | 212.7816 | 0.0004 |
| 15 | 212.7797 | 0.0012 | 212.7795 | 0.0013 |
| 16 | 212.7812 | 0.0010 | 212.7808 | 0.0009 |
| 17 | 212.7814 | 0.0017 | 212.7815 | 0.0017 |
| 18 | 212.7802 | 0.0013 | 212.7801 | 0.0012 |
| 19 | 212.7796 | 0.0009 | 212.7801 | 0.0010 |
| 20 | 212.7816 | 0.0017 | 212.7819 | 0.0012 |
| Mean pitch or Standard deviation of different locations | 212.7807 | 0.0011 | 212.7805 | 0.0011 |

To clarify the reason behind the picometer-level homogeneity, we conducted Cr grating pitch measurement at different sizes. Table 2 demonstrates evaluated mean pitch of Cr gratings structures measured at different sizes from 1 μm to 100 μm. It is obvious that the standard deviation will reach picometer level at 10 μm size. And for 1 μm size measurement including only 4 lines in the scan, the standard deviation is already as low as 0.11 nm. Since the averaged mean pitch begins to stay stable from 2 μm, which means the picometer-level homogeneity is mainly determined by the atom lithography technology process of a standing wave's "materialization" in the vacuum.

Table 2 Evaluated mean pitch of Cr gratings structures measured at different sizes

| Measurement | GC Method | | FT-FFT Method | |
|---|---|---|---|---|
| | Averaged mean pitch value, nm | Standard deviation, nm | Number of evaluation fields, nm | Standard deviation, nm |
| 100 μm x 10 μm | 212.7795 | 0.0005 | 212.7792 | 0.0006 |
| 50 μm x 10 μm | 212.7797 | 0.0005 | 212.7794 | 0.0005 |
| 20 μm x 10 μm | 212.7774 | 0.0018 | 212.7769 | 0.0018 |
| 10 μm x 10 μm | 212.7821 | 0.004 | 212.7828 | 0.007 |
| 5 μm x 5 μm | 212.7749 | 0.015 | 212.7870 | 0.016 |
| 2 μm x 2 μm | 212.783 | 0.06 | 212.802 | 0.11 |
| 1 μm x 1 μm | 212.88 | 0.11 | 212.6935 | 0.22 |

Table 3 Calibrated mean pitch of two different Cr gratings

| Sample | Calibrated mean pitch, nm |
|---|---|
| SN_1 | 212.782 ± 0.008 nm (*k*=2) |
| SN_2 | 212.781 ± 0.008 nm (*k*=2) |

As a working pitch standard in the length traceability chain, the Cr grating pitch should keep consistent at a very low uncertainty level from sample to sample. Therefore, we have calibrated randomly selected two different Cr gratings, and the calibrated mean pitches are 212.782 ± 0.008 nm (*k*=2) and 212.781 ± 0.008 nm (*k*=2). Since the calibrated results show that there is only 1 picometer discrepancy between two randomly selected samples, together with the same uncertainty of 0.008 nm (*k*=2), which means extreme consistency of Cr grating fabricated by atom lithography.

It must be emphasized that there are discrepancies among the theoretical pitch value (212.7705 ± 0.0049 nm), the measured pitch value based on diffraction method (212.7777 ± 0.0069 nm) and the measured pitch value based on metrological AFM (212.782 ± 0.008 nm) in this paper. Here are our considerations for these discrepancies. Firstly, it's very hard to evaluate which is the true value for Cr grating pitch. There are different error sources accumulation during the Cr grating fabrication process and the measurement process. Secondly, for the advanced manufacturing, the precise positioning repeatability and stability are much more important than the absolute positioning value. And meanwhile, picometer level discrepancies are acceptable for most cases. Thirdly, the Cr grating pitch's uncertainty level is comparable with the un-stabilized He-Ne laser. For the commonly used stabilized He-Ne laser, the wavelength is 632.9908nm with a fractional uncertainty[14] of $1.5 \times 10^{-6}$. For the Cr grating pitch, it has shown the mean pitch of 100 μm is 212.7795 nm with a standard deviation of 0.0005 nm (Table 2). In fact, for the grating interferometer application, the laser diffraction spot dimension is much bigger than 100 μm, which means the standard deviation will decrease further, also reaching a fractional uncertainty at $10^{-6}$ level. Therefore, based on the metrological measurement of Cr grating pitch for examination of accuracy, uniformity, consistency and uncertainty, we believe the Cr grating fabricated is acceptable to be used as a self-traceable pitch standard.

**Self-traceable angle standard.** Angular motion and displacement measurements are as important as linear for precision positioning[2], and the measurements are established by autocollimators, rotary encoders and multiple linear displacement sensors. Normally the arc length of the full circumference of a circle is the round angle of 2π rad (360°) can be used as a natural or self-traceable angle standard, the realization of an orthogonal angle at nanoscale is still in need. An orthogonal angle can not only be used as a transfer standard to calibrate the *xy*-plane of nearly all kinds of microscopes[43, 45], but also act as a key component in planar encoders for measuring 2D displacements or autocollimators[46, 47, 48]. Since we have fabricated the self-traceable length standard for the length traceability chain, here we aim at continue to fabricate a self-traceable angle standard of an orthogonal angle.

Like the full circumference of a circle can be used as a self-traceable 360° angle standard, the two diagonals of a rhombus are perpendicular, which can be used as a natural square ruler. Meanwhile, the rhombi array can be decomposed of two groups of one-dimensional parallel lines with same spacings. Since we have examined that the atom lithography can fabricate self-traceable pitch standard, we proposed a two-step atom lithography method to fabricate this natural square ruler[49], as illustrated in Fig.4. Based on the first single step atom lithography, a following atom lithography is conducted after rotating the substrate by θ (see Fig.4 (a)). In this way, the pattern created in the overlapping area is rhombus array of highly uniform dots (see Fig.4 (b)). Because the natural square ruler features formed in the rhombus dot array which is independent of the rotation angle, so the orthogonal diagonals inside a two-dimensional Cr nano-grating can be used as a self-traceability angle standard.

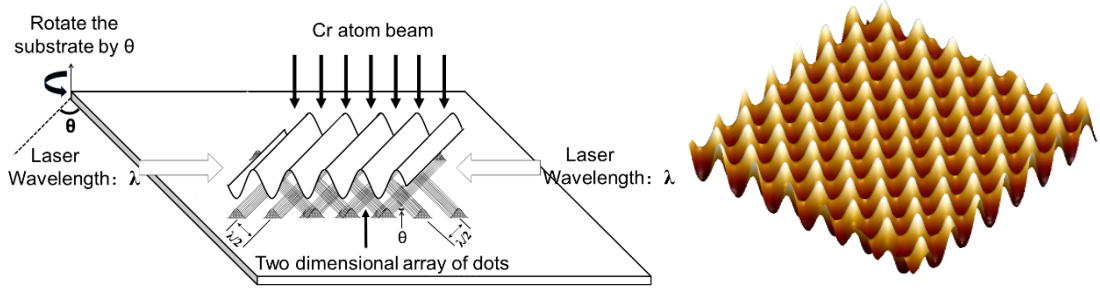

Fig.4 (a) Schematic of two-step laser focused atomic deposition; (b)3D AFM image (2 μm×2 μm) of the rhombus pattern of two-step atom lithography[49].

Similarly, we also conducted the metrological AFM measurements to examine the accuracy, uniformity and consistency of self-traceable angle with the Met.LR-SPM at PTB. The metrological measurements were conducted in the PTB cleanroom center, and the temperature of the Met.LR-SPM and surrounding air was 21.0±0.5 ℃, the relative humidity of the air was 44±2 %. Two groups of measurement with two different area size are measured: 100 μm×1 μm with 50000 pixels/line×20 lines; 20 μm×1 μm with 50000 pixels/line×20 lines. The measurements were repeated 5 times each and different locations were measured. Table 4 shows the metrological measurement results of the orthogonal angle in randomly selected two-dimensional Cr nano-gratings. As expected, the two-dimensional Cr nano-gratings show extremely accurate and uniform orthogonal angle, with final calibrated results: 90.001°± 0.003° ($k$=2) and 90.000°± 0.003° ($k$=2). The discrepancy between the orthogonal angle for two samples is only 0.001° and the expanded uncertainty is only 0.003°. It agrees well with previous theoretical uncertainty budget calculation of 0.0027°[49].

Table 4 Calibrated orthogonal angle results of two-dimensional Cr nano-gratings

| Measurement Filed | Sample 1 | | Sample 2 | |
|---|---|---|---|---|
| | Orthogonal angle | Standard deviation | Orthogonal angle | Standard deviation |
| MF1 | 90.0004° | 0.0013° | 89.9991° | 0.0003 |
| MF2 | 90.0006° | 0.0007° | 89.9996° | 0.0002 |
| MF3 | 90.0007° | 0.0002° | 90.0000° | 0.0004 |
| Mean angle | 90.0006° | | 89.9996° | |
| $\sigma_{uniformity}$ | 0.0001° | | 0.0005° | |
| $\sigma_{repeatability}$ | 0.0013° | | 0.0004° | |
| Final calibration results | 90.001°± 0.003° ($k$=2) | | 90.000°± 0.003° ($k$=2) | |

The extremely high accuracy, uniformity and consistency of the orthogonal angle inside the two-dimensional Cr grating in fact reflects the extremely high accuracy, uniformity and consistency of the atom lithography. Because we transfer the length standard into angle standard, the accuracy of length standard determines the latter. Besides, the rotation angle has no effect on the self-traceable angle standard which offers great convenience for us to fabricate an ideal natural square ruler randomly. Therefore, we believe the two-dimensional Cr grating fabricated is acceptable to be used as a self-traceable angle standard.

**Self-traceable grating interferometer.** After the self-traceable length grating standard has been achieved, we continue to build a self-traceable grating interferometer as the dynamic displacement measurement method for the newly proposed length traceability chain. The self-traceable grating interferometer displacement measurement system adopts a zero-difference detection scheme and some key rules are followed. To lower the influence of the external environment on the system, two interferometric optical paths follow the same path and keep the optical range difference between the two beams as zero as possible. And the polarization characteristics of light and optical elements are fully applied to improve the efficiency of light and achieve the direction discrimination of displacement measurement based on four orthogonal sinusoidal signals with a phase difference of 90°. Therefore, we designed the self-traceable grating interferometer system in Fig.5(a) and established the experimental setup under cage style configurations in Fig.5(b). The polarized light from the single-frequency laser at 375 nm is uniformly split into two beams, R1 and R2, and diffracted onto the self-traceable grating surface under Littrow conditions. Then $+1^{st}$ order diffracted light from R1 and $-1^{st}$ order diffracted light from R2 are combined by the polarization system and interfered to form interference fringes with a 90° phase shift, which are detected by photodetectors PD1 to PD4. And the moving stage to be measured moves along the grating vector direction.

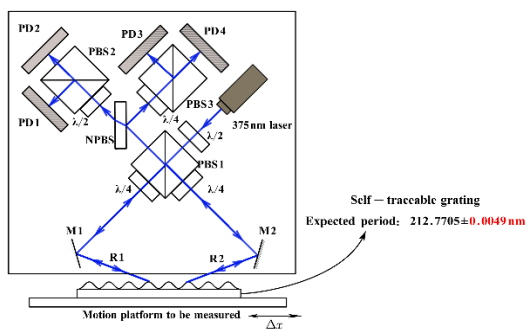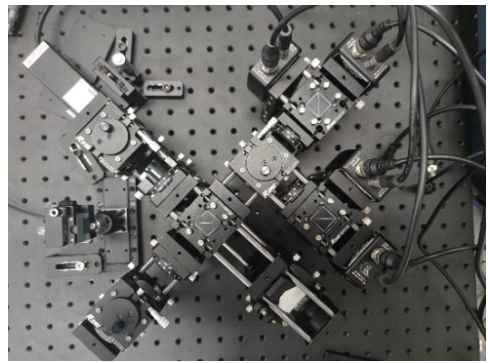

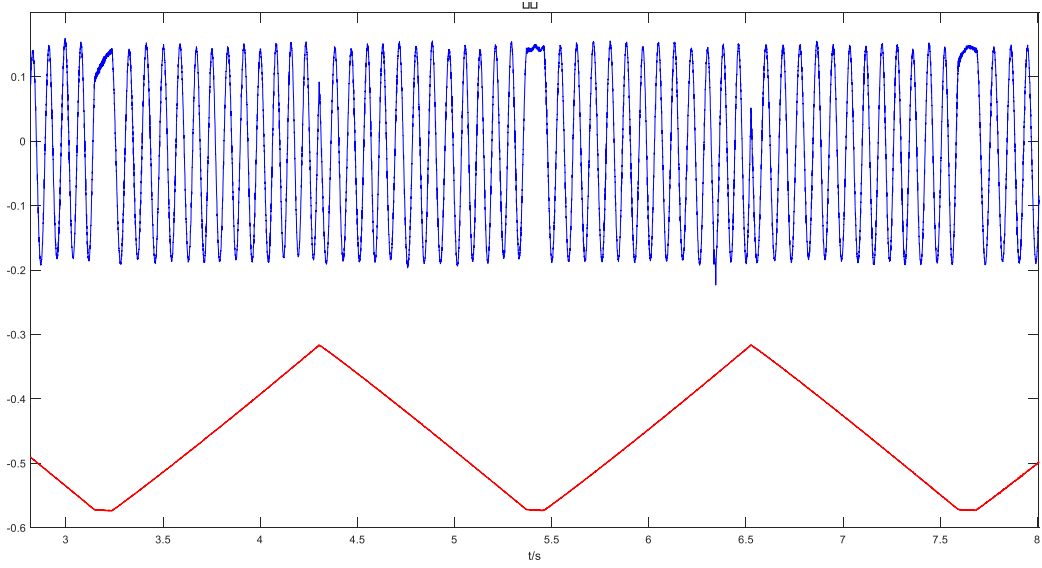

Fig.5 (a) Optical design of self-traceable grating interferometer system; (b)Experimental setup of self-traceable grating interferometer system; (c) Signal detected by photodetectors under a triangle motion of the stage.

The optical trajectory of R1 and R2 are described as follows[22, 50, 51]:

$$\begin{aligned}\overrightarrow{E_{R1}} &= J_{NPBS} J_{PBS1\cdot S} J_{QW(\pi/4)} J_{M_1} J_{G1} J_{M_1} J_{QW(\pi/4)} J_{PBS1\cdot P} J_{HW} \overrightarrow{E_L} \\ \overrightarrow{E_{R2}} &= J_{NPBS} J_{PBS1\cdot P} J_{QW(\pi/4)} J_{M_2} J_{G2} J_{M_2} J_{QW(\pi/4)} J_{PBS1\cdot S} J_{HW} \overrightarrow{E_L}\end{aligned} \quad (1)$$

Where $J_G = \begin{bmatrix} -\sqrt{\eta_p} & 0 \\ 0 & \sqrt{\eta_s} \end{bmatrix} e^{i\phi_D}$ is Jones matrix for the self-traceable gratings, and $\eta_p$ and $\eta_s$ are the ±1st order diffraction efficiency of self-traceable gratings under TM and TE polarization conditions, and $\phi_D$ is Doppler frequency shift due to the motion of the displacement stage following the equation:

$$\begin{cases} \phi_{D1} = +\dfrac{2\pi}{d}\Delta x \\ \phi_{D2} = -\dfrac{2\pi}{d}\Delta x \end{cases} \quad (2)$$

Where $\Delta x$ and $d$ are distance travelled by the motion stage to be measured in the direction of the grating vector and the self-traceable grating pitch. And for the polarized beam detection section, the light field intensity superimposed at the photodetectors PD1 to PD4 are:

$$\begin{cases} \overrightarrow{E_{\text{PD1}}} = J_{\text{PBS2·S}} J_{\text{HW}(3\pi/8)} \left( \overrightarrow{E_{\text{R1}}} + \overrightarrow{E_{\text{R2}}} \right) \\ \overrightarrow{E_{\text{PD2}}} = J_{\text{PBS2·P}} J_{\text{HW}(3\pi/8)} \left( \overrightarrow{E_{\text{R1}}} + \overrightarrow{E_{\text{R2}}} \right) \\ \overrightarrow{E_{\text{PD3}}} = J_{\text{PBS3·S}} J_{\text{QW}(\pi/4)} \left( \overrightarrow{E_{\text{R1}}} + \overrightarrow{E_{\text{R2}}} \right) \\ \overrightarrow{E_{\text{PD4}}} = J_{\text{PBS3·P}} J_{\text{QW}(\pi/4)} \left( \overrightarrow{E_{\text{R1}}} + \overrightarrow{E_{\text{R2}}} \right) \end{cases} \quad (3)$$

According to $I_{\text{PD}} = \left\| \overrightarrow{E_{\text{PD}}} \right\|$, the photodetector receives interference light with an intensity as follows:

$$\begin{cases} I_{\text{PD1}} \propto I_0 \left[ \cos\left( \frac{4\pi}{d} \Delta x + \frac{\pi}{2} \right) + 1 \right] \\ I_{\text{PD2}} \propto I_0 \left[ \cos\left( \frac{4\pi}{d} \Delta x - \frac{\pi}{2} \right) + 1 \right] \\ I_{\text{PD3}} \propto I_0 \left[ \cos\left( \frac{4\pi}{d} \Delta x \right) + 1 \right] \\ I_{\text{PD4}} \propto I_0 \left[ \cos\left( \frac{4\pi}{d} \Delta x - \frac{\pi}{2} \right) + 1 \right] \end{cases} \quad (4)$$

For every $d/2$ movement of the moving platform, the interferometric fringe changes by one complete sinusoidal cycle, and the displacement direction discrimination is achieved by judging the relationship of the phases from PD1 to PD4. The four signals are differentially processed to obtain two orthogonal signals, and the phase values over time are obtained by the inverse tangent method and the stage displacement $\Delta x$ are as follows:

$$\theta(\Delta t) = \tan^{-1} \left[ \frac{\phi_a(\Delta t)}{\phi_b(\Delta t)} \right] \quad (5)$$

$$x(\Delta t) = \tan^{-1} \left[ \frac{\phi_a(\Delta t)}{\phi_b(\Delta t)} \right] \times \frac{106.38525}{2\pi} \quad (6)$$

Based on the above measurement principle, the measurement scale graduation is half of the self-traceable grating pitch, which is determined by the Cr atom transition frequency originally. Here we set the theoretical pitch value of 212.7705 nm as the standard value and the unit of equation (6) is nanometer. Therefore, compared to other grating interferometers, the self-traceable grating interferometer will be the first time capable to achieve direct traceability of grating interferometry in terms of physical mechanism. In Fig.5(c), a piezoelectric ceramic stage is driven by a triangular wave and moves at a uniform reciprocating speed, and the blue line is the curve of interference light intensity as a function of time received by a single photodetector and the red line is the measured displacement of the system solved according to equation

(6). As can be seen from the graph, the displacement measurements of the system perfectly reproduce the triangular wave uniform motion characteristics of the piezoelectric ceramic stage, which can initially prove the feasibility of this experimental system and displacement algorithm. Here it should be emphasized that besides self-traceability, with an extreme high line density of 4700 lines/mm, a raw displacement signal of 106.4 nm can be obtained with only twice the optical subdivision. Followed by subdivision of electro-optical signals, the resolution of displacement measurement will be greatly improved.

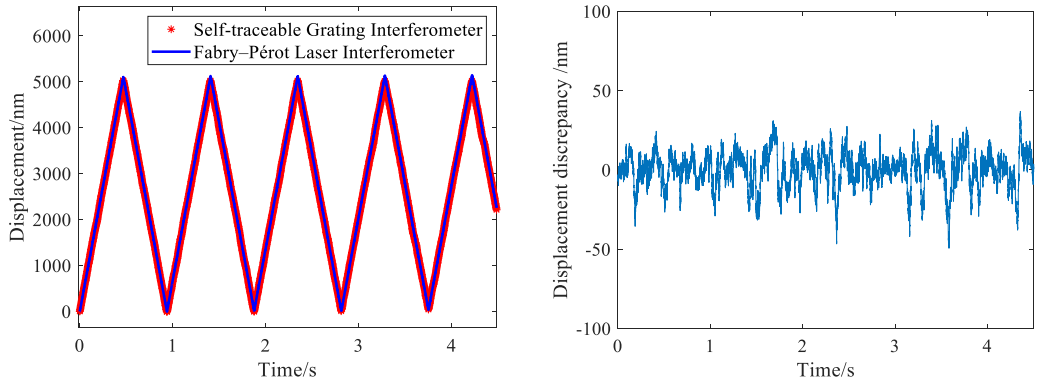

Fig.6 (a) Displacement of a moving stage measured by self-traceable grating interferometer and commercial laser interferometer; (b) Displacement measurement discrepancy of self-traceable grating interferometer and commercial laser interferometer in Fig.6(a).

To test the measurement performance of self-traceable grating interferometer, we conduct a comparison measurement with a commercial laser interferometer (quDIS, Fabry–Pérot interferometer). The laser interferometer reflector and the self-traceable grating are mounted simultaneously on different planes of the moving stage to be measured, with the reflector mounted perpendicular to the forward direction of the moving stage and the self-traceable grating mounted parallel to the direction of motion of the moving stage. The nominal moving range is set at 5000 nm and the stage is driven by a triangular wave at a frequency of 1Hz, and the data sampling number is 25000. Fig.6 shows the displacement measurement of self-traceable grating interferometer and the Fabry–Pérot laser interferometer and the corresponding discrepancy. After Abbe error correction, it's obvious that the self-traceable gratings grating interferometer shows dynamic triangular motion characteristics of the stage which is comparable with the commercial laser interferometer. The displacement discrepancy fluctuates mainly between ±20 nm, which is 0.4% of the whole motion range, indicating good consistency between the self-traceable grating interferometer and the laser interferometer. We speculate that he discrepancy can be further decreased by diffraction efficiency improvement and better noise suppression. Therefore, we believe the self-traceable grating interferometer based on Cr atom lithography holds independent traceability to meter for dynamic displacement measurement.

## Discussion

We demonstrate self-traceable length standard, self-traceable angle standard and self-traceable grating interferometer based on Cr atom lithography, and metrological comparison measurements with laser interferometry have proved their extremely high accuracy, uniformity and consistency. The newly proposed length traceability chain is promising to bring key improvement for precision positioning technology and scientific instrument development. Here we give some potential predictions.

AFM is a versatile tool for nanometer scale measurement. Because the AFM scanner based on the piezoelectric ceramics has the disadvantage of non-linearity and angle aberrations, the AFM should be calibrated by transfer length standards periodically. The self-traceable length standard and angle standard based on Cr atom lithography have extremely high accuracy, which is ideal for the AFM calibration. Furthermore, the metrological AFM is the most accurate AFM which has position measurement with integrated laser interferometers. Due to the strict requirement of the laser wavelength stability, metrological AFMs have only been used in National Metrology Institutes. The self-traceable grating interferometer is easy to be designed as a compact sensor with direct traceability to meter at low cost. Therefore, a new type of metrological AFM equipped with self-traceable grating interferometers can be developed for the industry. The widely used metrological AFMs will improve the accuracy for the customers.

Accelerometer is a key device for measuring the linear acceleration of a carrier, and its measurement accuracy, consistency and reliability are the common technical requirements of acceleration applications. The essence of the acceleration measurement is the ultra-precision displacement measurement and calibration technology, and the traditional accelerometer calibration method based on the laser interferometer is more and more difficult to meet the urgent needs of the flattening of the value transfer. To meet the needs of developing traceable, ultra-precise, on-a-chip, embedded, and directly calibrated accelerometers, a micro-optic-electro-mechanical accelerometer based on self-traceable grating interferometers can be a proper solution. And the relative uncertainty of *in situ* calibration of the accelerometer is increased to the order of $1\times10^{-6}$, which promotes the flattening of the value transfer in the field of acceleration measurement.

Compared with other types grating interferometers, the self-traceable grating interferometers holds a direct traceability to meter, which simplifies the calibration process. And the grating line density reaches 4700 lines per mm, which improves the displacement measurement resolution dramatically. And for the two-dimensional self-traceable grating interferometers, the self-traceable angle standards will offer direct angle traceability and extremely high accurate orthogonal angle, which will help decrease the linear error in multi-axis systems. Therefore, it can be expected that self-traceable interferometers will play an important role in the field of nano-positioning.